\documentclass[11pt]{article}

\usepackage{amssymb}
\usepackage{graphics}

\newtheorem{thm}{Theorem}
\newtheorem{clm}{Claim}[thm]
\newtheorem{theorem}{Theorem}[section]

\newtheorem{lemma}[theorem]{Lemma}

\def \no {\noindent}
 \def \sm {\setminus}

\title{Weighted Independent Sets in a\\ Subclass of $P_6$-free Graphs}

\author{T. Karthick\footnote{Computer Science Unit, Indian Statistical
Institute, Chennai Centre, Chennai-600113, India. E-mail: karthick@isichennai.res.in}}

\begin{document}
\maketitle

\begin{abstract}
The {\it Maximum Weight Independent Set (MWIS)} problem on graphs with vertex weights
asks for a set of pairwise nonadjacent vertices of maximum total weight. The complexity of
the MWIS problem for $P_6$-free graphs is unknown.

In this note, we show that the  MWIS problem  can be solved in time $O(n^3m)$ for ($P_6$, banner)-free graphs by analyzing the structure of subclasses of these class of graphs.  This extends the existing results for ($P_5$, banner)-free graphs, and ($P_6$, $C_4$)-free graphs. Here, $P_t$ denotes the chordless path on $t$ vertices, and a {\it banner} is the graph obtained from a  chordless cycle on four vertices by adding a vertex that has exactly one neighbor on
the cycle.

\end{abstract}

\no{\bf Keywords}: Graph algorithms; Independent sets;  $P_6$-free graphs.

\section{Introduction}

In an undirected graph $G$, an {\it independent set} is a set of mutually nonadjacent vertices.
 The \textsc{Maximum Weight Independent Set (MWIS)} problem
asks for an independent set of maximum total  weight in the given graph $G$ with vertex weight function $w$ on  $V(G)$.
 The \textsc{Maximum Independent Set (MIS)}  problem is the MWIS problem where all the vertices $v$ in $G$ have the same weight $w(v)=1$.
 The MWIS problem on graphs ([GT20] in \cite{GJ}) is one of the most investigated problems on graphs because of its applications in computer science,
 operations research, bioinformatics and other fields, including train dispatching \cite{FMSWZ} and data
mining \cite{TYT}.

If $\cal{F}$ is a family of graphs, a graph
 $G$ is said to be {\it $\cal{F}$-free} if it contains no
induced subgraph isomorphic to any graph in $\cal{F}$.

The MWIS problem is  known to be $NP$-complete in general and
hard to approximate; it remains $NP$-complete even on restricted classes of graphs such as triangle-free graphs \cite{Poljak}, and
($K_{1, 4}$,diamond)-free graphs \cite{Corneil}. Alekseev \cite{Alek} showed that the M(W)IS problem remains $NP$-complete on $H$-free graphs, whenever $H$ is connected, but neither a path nor a subdivision of the claw ($K_{1, 3}$).
On the other hand, the MWIS problem is known to be solvable in polynomial time
on many graph classes, such as chordal graphs \cite{Frank}, perfect graphs \cite{GLS}, $2K_2$-free graphs \cite{Farber}, claw-free graphs \cite{Minty}, and fork-free graphs \cite{LM}. It is well known that the MWIS problem is solvable in linear time for the class of $P_4$-free graphs (also known as co-graphs) \cite{CPS}.

In this note, we focus on graphs which do not contain certain induced paths.
 As a natural generalization of $P_4$-free graphs, the class of $P_k$-free graphs ($k \geq 5$) has been studied widely in the literature. The complexity of the MWIS problem for
$P_5$-free graphs was unknown for several decades.  Recently, Lokshantov, Vatshelle
and Villanger \cite{LVV} showed an $O(n^{12}m)$ time algorithm for the MWIS problem on
$P_5$-free graphs via minimal triangulations.  However, the complexity of the MWIS problem is
unknown for the class of $P_6$-free graphs.
The MWIS problem is known to be solvable efficiently  on several subclasses of $P_k$-free graphs ($k \geq 5$) by several techniques, and we refer to \cite{BH,  GL, K} and the references therein for a survey. %Among these techniques, graph decomposition techniques received much attention recently \cite{MSK, BH, BGM, BKLM,  BLM,  K, LM}.

A vertex $z \in V(G)$ {\it distinguishes} two other vertices $x, y \in V(G)$ if $z$ is adjacent to one of them and
nonadjacent to the other. A vertex set $M\subseteq V(G)$ is a {\it module} in $G$ if no vertex from $V(G) \setminus M$ distinguishes
two vertices from  $M$. The {\it trivial modules} in $G$ are $V(G)$, $\emptyset$, and all one-elementary vertex sets.
A graph $G$ is {\it prime} if it contains only trivial modules. Note that prime graphs with at least three vertices are connected.

A class of graphs $\cal{G}$ is {\it hereditary} if every induced
subgraph of a member of $\cal{G}$ is also in $\cal{G}$.  We will use
the following theorem by L\"ozin and Milani\v{c} \cite{LM}.

\begin{thm}[\cite{LM}]\label{thm:LM}
Let $\cal{G}$ be a hereditary class of graphs.  If the MWIS problem
can be solved in $O(n^p)$-time for prime graphs in $\cal{G}$, where $p
\geq 1$ is a constant, then the MWIS problem can be solved for graphs
in $\cal{G}$ in time $O(n^p + m)$.  \hfill{\small $\blacksquare$}
\end{thm}

A {\it clique} in $G$ is a subset of pairwise adjacent vertices in $G$.
A {\it clique separator/clique cutset} in a connected graph $G$ is a subset $Q$ of vertices in $G$ such that $Q$ is a clique and
such that the graph induced by $V(G) \sm Q$ is disconnected. A graph
is an {\it atom} if it does not contain a clique separator.

Let $\cal{C}$ be a class of graphs.  A graph $G$ is {\it nearly
$\cal{C}$} if for every vertex $v$ in $V(G)$ the graph induced by
$V(G) \setminus N[v]$ is in $\cal{C}$.  We will also use the following theorem given in  \cite{MSK}. Though the theorem (Theorem 1 of \cite{MSK}) is stated only for hereditary class of graphs, the proof also work for any class of graphs, and is given below:

\begin{thm}[\cite{MSK}]\label{thm:Tar}
Let $\cal C$ be a class of graphs such that MWIS can be solved in time
$O(f(n))$ for every graph in $\cal C$ with $n$ vertices.  Then in any
class of graphs whose atoms are all nearly $\cal C$ the
MWIS problem can be solved in time $O(n\cdot f(n)+nm)$.  \hfill {\small $\blacksquare$}
\end{thm}

We see that the Theorems \ref{thm:LM} and \ref{thm:Tar} can be combined as follows:

\begin{thm}\label{prime-atoms-comb}
Let $\cal{G}$ be a hereditary class of graphs. Let $\cal{P}$ denotes the class of prime graphs in $\cal{G}$. Let $\cal C$ be a class of graphs such that MWIS can be solved in time $O(f(n))$ for every graph in $\cal C$ with $n$ vertices. Suppose that every atom of a graph $G \in \cal{P}$ is nearly $\cal C$. Then the
MWIS problem in $\cal{G}$ can be solved in time $O(n\cdot f(n)+nm)$.

\end{thm}

\no{\bf Proof}. Let $G$ be a graph in $\cal G$.  First suppose that $G \in \cal P$. Since every atom of $G$
is nearly $\cal C$, and since the MWIS problem for graphs in $\cal C$
can be solved in time $O(f(n))$, MWIS can be solved in
time $O(n \cdot f(n)+nm)$ for $G$, by Theorem~\ref{thm:Tar}.  Then the time
complexity is the same when $G$ is not prime, by Theorem
\ref{thm:LM}.
\hfill {\small $\blacksquare$}

\medskip
In this note, using the above framework, we show that the MWIS problem in ($P_6$, banner)-free graphs can be solved in time $O(n^3m)$, by analyzing the atomic
structure and the MWIS problem in various subclasses of ($P_6$, banner)-free graphs, where a {\it banner} is the graph obtained from a  chordless cycle on four vertices by adding a vertex that has exactly one neighbor on
the cycle (see also Figure~\ref{sg}). This result extends the results known for $P_4$-free graphs,  ($P_6, C_4$)-free graphs, and for ($P_5$, banner)-free graphs \cite{BH, L}.

We note that applying Corollary 9 in \cite{BH} which used an approach for solving MWIS by combining prime graphs and atoms, it was claimed in \cite{BKLM} that MWIS is solvable efficiently in time $O(n^7m)$ for ($P_6$, banner)-free graphs. However, Corollary 9 in \cite{BH} is not proven (and thus has to be avoided).

It is also noteworthy that the MWIS problem remains $NP$-complete in
banner-free graphs (this follows from the result of Murphy \cite{Murphy} for graphs with large girth). The class of
banner-free graphs is of particular interest, since it
contains two important subclasses where the MWIS problem can be solved efficiently, namely
claw-free graphs and $P_4$-free graphs. Also, the complexity of the MIS Problem for various subclasses of
banner-free graphs has been studied in the literature, and we refer to \cite{GHL, L} for more details.

\section{Notation and Terminology}\label{NT}

\begin{figure}[t]
\centering
 \includegraphics{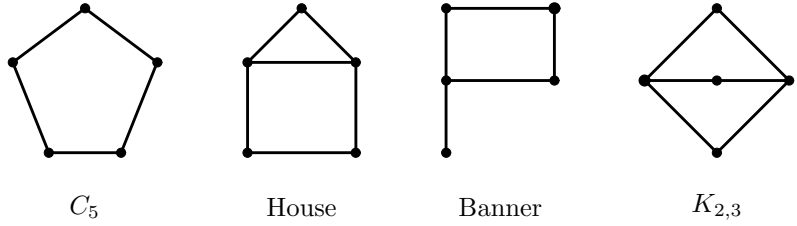}
\caption{Some special graphs}
\label{sg}
\end{figure}

For notation and terminology not defined here, we follow \cite{BLeS}. Let $G$ be a finite, undirected, simple graph with vertex-set
$V(G)$ and edge-set $E(G)$.  We let $|V(G)| = n$ and $|E(G)| = m$. The symbols
$P_k$ and $C_k$ respectively denotes the chordless path and the chordless cycle
on $k$ vertices. Let $K_{m, n}$ denote the complete bipartite graph with $m$ vertices in one partition set and $n$ vertices in the other. The {\it banner} is also called $P$ or {\it $4$-apple} or $A_4$ in various papers \cite{BH, BKLM, BLM}, and see Figure \ref{sg} for some of the special graphs that we have used in this paper.

  For a
vertex $v \in V(G)$, the {\it neighborhood} $N(v)$ of $v$ is the set
$\{u \in V(G) \mid uv \in E(G)\}$, and its {\it closed neighborhood}
$N[v]$ is the set $N(v) \cup \{v\}$.  The neighborhood $N(X)$ of a
subset $X \subseteq V(G)$ is the set $\{u \in V(G)\setminus X  \mid  u$ $
\mbox{~is adjacent to a vertex of }X\}$, and its closed neighborhood
$N[X]$ is the set $N(X)\cup X$. Let $\overline{N[X]}$ denote the set $V(G) \sm N[X]$. Given a subgraph $H$ of $G$ and $v
\in V(G)\setminus V(H)$, let $N_H(v)$ denote the set $N(v) \cap V(H)$,
and for $X \subseteq V(G)\setminus V(H)$, let $N_H(X)$ denote the set
$N(X) \cap V(H)$.  For any two subsets $S$, $T\subseteq V(G)$, we say
that $S$ is \emph{complete} to $T$ if every vertex in $S$ is adjacent
to every vertex in $T$.

The following notation will be used several times in the proofs.
Given a graph $G$, let $v$ be a vertex in $G$ and $H$ be an induced
subgraph of $G\setminus N[v]$.  Let $t=|V(H)|$.  Then we define the
following sets:
\begin{eqnarray*}
Q_{\ } &=& \mbox{the component of $G\setminus N[H]$ that contains
$v$}, \\
A_i &=& \{x \in V(G) \setminus V(H) \mid |N_H(x)| = i\} \ (1 \leq i
\leq t), \\
A_i^+ &=& \{x \in A_i \mid N(x) \cap Q \neq \emptyset\}, \\
A^-_i &=& \{x \in A_i \mid N(x) \cap Q = \emptyset\}, \\
A^+ &=&   A^+_1\cup\cdots\cup A^+_t \ \mbox{and} \
A^- =  A^-_1\cup\cdots\cup A^-_t.
\end{eqnarray*}
So, $N(H) = A^+ \cup A^-$.  Note that, by the definition of $Q$ and
$A^+$, we have $A^+ = N(Q)$.  Hence $A^+$ is a separator between $H$
and $Q$ in $G$.

\section{MWIS in ($P_6$, banner)-free graphs}

In this section, we prove that the  MWIS problem in ($P_6$, banner)-free graphs can be solved in $O(n^3m)$-time,  by analyzing the atomic
structure and the MWIS problem in various subclasses of ($P_6$, banner)-free graphs.

\subsection{MWIS  in ($P_6$, $C_4$)-free graphs}

\begin{thm} \label{p6-c4-free-complexity}
 The MWIS problem can be solved in time $O(nm)$ for ($P_6$, $C_4$)-free graphs.
\end{thm}

\no{\bf Proof}. In \cite{BH}, Brandst\"{a}dt and Ho\'{a}ng showed that atoms of ($P_6, C_4$)-free graphs
are either nearly chordal or 2-specific graphs (see \cite{BH} for the definition of $2$-specific graphs). Since the MWIS
problem is trivial for 2-specific graphs, and can be solved in time $O(m)$ for  chordal graphs \cite{Frank}, by Theorem \ref{thm:Tar},
the MWIS problem for ($P_6, C_4$)-free graphs can be solved in time $O(nm)$.  \hfill{\small $\blacksquare$}

\subsection{MWIS  in ($P_6$, banner, house)-free graphs}

In this section, we will show that the MWIS can be solved in $O(nm)$-time for ($P_6$, banner, house)-free graphs.
Though the following lemma can be derived from a result of Ho\'{a}ng and Reed \cite{HR}, we give a simple proof here  for completeness.

\begin{lemma} \label{prime-HB-free}
If $G = (V, E)$ is a prime (banner, house)-free graph, then $G$ is $C_4$-free.
\end{lemma}

\no{\bf Proof}. Suppose to the contrary that $G$ contains an induced $C_4$ with vertex set $\{a_1, a_2, b_1, b_2\}$ and edge set
 $\{a_1b_1, b_1a_2, a_2b_2, b_2a_1\}$. Let $Q$ be the connected
component in the complement of the graph $G[N(a_1) \cap N(a_2)]$ that contains
$b_1$ and $b_2$. Since $G$ is prime, there exist vertices $b_1', b_2' \in Q$ such that
$b_1'b_2' \notin E$, which are distinguished by a vertex $z \notin Q$, say $zb_1' \in E$
and $zb_2' \notin E$. Then since $\{z, b_1', b_2', a_1, a_2\}$ does not induce a house or a banner in $G$, we have $za_1 \in E$ and $za_2 \in E$. Hence $z \in Q$, which is a contradiction. This shows Lemma \ref{prime-HB-free}. \hfill{\small $\blacksquare$}

\medskip
Then we immediately have the following:

\begin{thm} \label{p6-house-banner-free-complexity}
 The MWIS problem can be solved in $O(nm)$-time for ($P_6$, banner, house)-free graphs.
\end{thm}

\no{\bf Proof}. Let $G$ be a prime ($P_6$, banner, house)-free graph. Then by Lemma \ref{prime-HB-free}, $G$ is ($P_6$, $C_4$)-free. Since the
MWIS problem  can be solved in time $O(nm)$ for ($P_6$, $C_4$)-free graphs (by Theorem \ref{p6-c4-free-complexity}),  the MWIS problem can be solved in time $O(nm)$ for prime ($P_6$, banner, house)-free graphs. Then the time complexity is same when $G$ is not prime, by Theorem \ref{thm:LM}.
\hfill{\small $\blacksquare$}

\subsection{MWIS problem in ($P_6$, banner, $C_5$)-free graphs}

In this section, we show that the MWIS problem can be solved in time $O(n^2m)$ for ($P_6$, banner, $C_5$)-free graphs.
We use the following
lemma given by Brandst\"{a}dt et al. in \cite{BKLM}.

\begin{lemma}[\cite{BKLM}] (see also \cite{BLM}) \label{banner-free--k23-free}
Prime banner-free graphs are $K_{2, 3}$-free. \hfill{\small $\blacksquare$}
\end{lemma}

\begin{thm} \label{prime-p6-Banner-c5-free-implies-house-free}
Let $G$ be a prime ($P_6$, banner, $C_5$)-free graph. Then every
atom of $G$ is nearly house-free.
\end{thm}

\noindent{\it Proof.} Let $G'$ be an atom of $G$.  We want to show
that $G'$ is nearly house-free, so let us assume on the contrary
that there is a vertex $v \in V(G')$ such that $G'\setminus N[v]$
contains an induced house $H$.  Let $H$ have vertex set $\{v_1, v_2,
v_3, v_4, v_5\}$ and edge set $\{v_1v_2, v_2v_3, v_3v_4, v_4v_1, v_2v_5, v_3v_5\}$.  For $i=1, 2, \ldots, 5$,  we define sets $A_i$,
$A_i^+$, $A_i^-$, $A^+$, $A^-$, and $Q$ as in the last paragraph of
Section~2.

Note that by the definition of $Q$ and $A^+$, we have $A^+ = N(Q)$.  Hence $A^+$ is a separator between $H$ and $Q$ in $G$.  Now we have the following:

\begin{clm}\label{cl:a1p} $A_1^+ = \emptyset$.\end{clm}

\no{\it Proof of Claim ~\ref{cl:a1p}}. Suppose not, and let $x \in A_1^+$.  Then by the definition of $A_1^+$, there exists a vertex $y \in Q$ such that $xy \in E$. Now:
 \begin{enumerate}
\item[(i)] If $N_H(x) = \{v_1\}$ or $\{v_2\}$, then $\{v_1, v_2, v_3, v_4, x\}$ induces a banner in $G$, which is a contradiction.
\item[(ii)] If  $N_H(x) = \{v_5\}$, then $\{y, x, v_5, v_3, v_4, v_1\}$ induces a $P_6$ in $G$, which is a contradiction.
 \end{enumerate}
 Since the other cases are symmetric, Claim \ref{cl:a1p} is proved. $\blacklozenge$

\begin{clm}\label{cl:a2p} If $x \in A_2^+$, then $N_H(x)$ is either $\{v_2, v_3\}$ or $\{v_1, v_4\}$. \end{clm}

\no{\it Proof of Claim~\ref{cl:a2p}}. Suppose not. Since $x \in A_2^+$, there exists a vertex $y \in Q$ such that $xy \in E$. Now:
\begin{enumerate}
\item[(i)] If $N_H(x) = \{v_1, v_2\}$, then  $\{y, x, v_1, v_4, v_3, v_5\}$ induces a $P_6$ in $G$, which is a contradiction.
 \item[(ii)] If $N_H(x) = \{v_2, v_4\}$, then  $\{y, x, v_2, v_3, v_4\}$ induces a banner in $G$, which is a contradiction.
 \item[(iii)] If $N_H(x) = \{v_1, v_5\}$, then  $\{v_1, v_2, v_5, x, y\}$ induces a banner in $G$, which is a contradiction.
 \item[(iv)] If $N_H(x) = \{v_2, v_5\}$, then  $\{y, x, v_5, v_3, v_4, v_1\}$ induces a $P_6$ in $G$, which is a contradiction.
 \end{enumerate}
Since the other cases are symmetric, Claim \ref{cl:a2p} is proved. $\blacklozenge$

\medskip

By Claim \ref{cl:a2p}, we define sets $B_{1} = \{x \in A_2^+ \mid N_H(x) = \{v_2, v_3\}\}$ and $B_{2} = \{x \in A_2^+ \mid N_H(x) = \{v_1, v_4\}\}$. Then $A_2^+ = B_{1} \cup B_{2}$.

\medskip

\begin{clm}\label{cl:B}  If $B_{1} \neq \emptyset$, then $B_{2} = \emptyset$, and vice versa. \end{clm}

\no{\it Proof of Claim~\ref{cl:B}}. Assume the contrary, and let $x \in B_{1}$ and $y \in B_{2}$. Then since $\{x, y, v_1, v_2, v_5\}$ does not induce a banner in $G$, $xy \notin E$. Since $x, y \in A_2^+$ and $Q$ is connected, there exists a path $z_0-\cdot\cdot\cdot-z_p$ inside $Q$ such that $xz_0 \in E$ and $yz_p \in E$, and we choose a shortest such path. If $p = 0$, then  $\{z_0, x, v_2, v_1, y\}$ induces a $C_5$ in $G$, which is a contradiction. So, $p \geq 1$. But, then $\{z_1, z_0, x, v_2, v_1, v_4\}$ induces a $P_6$ in $G$, which is a contradiction. So the claim holds. $\blacklozenge$

\begin{clm}\label{cl:a3p}   If $x \in A_3^+$, then $N_H(x) = \{v_2, v_3, v_5\}$. \end{clm}

\no{\it Proof of Claim~\ref{cl:a3p}}. Suppose not. Since $x \in A_3^+$, there exists a vertex $y \in Q$ such that $xy \in E$. Now:
\begin{enumerate}
\item[(i)] If $N_H(x)= \{v_1, v_2, v_3\}$, then $\{y, x, v_1, v_4, v_3\}$ induces a banner in $G$, which is a contradiction.

\item[(ii)] If $N_H(x) = \{v_1, v_2, v_4\}$, then   $\{y, x, v_2, v_3, v_4\}$ induces a banner in $G$, which is a contradiction.

\item[(iii)] if $N_H(x) = \{v_1, v_4, v_5\}$ or  $\{v_1, v_3, v_5\}$, then  $\{y, x, v_1, v_2, v_5\}$ induces a banner in $G$, which is a contradiction.

\item[(iv)] if $N_H(x) = \{v_1, v_2, v_5\}$, then  $\{x, v_1, v_4, v_3, v_5\}$ induces a $C_5$ in $G$, which is a contradiction.
\end{enumerate}
Since the other cases are symmetric, Claim \ref{cl:a3p} is proved. $\blacklozenge$

\begin{clm}\label{cl:a4p} If $x \in A_4^+$, then  $N_H(x) = \{v_1, v_{2}, v_{3}, v_{4}\}$. \end{clm}

\no{\it Proof of Claim~\ref{cl:a4p}}. Suppose not. Since $x \in A_4^+$, there exists a vertex $y \in Q$ such that $xy \in E$. Now:
\begin{enumerate}
\item[(i)] If $N_H(x) = \{v_1, v_2, v_3, v_5\}$, then  $\{y, x, v_1, v_4, v_3\}$  induces a
banner in $G$, which is a contradiction.

\item[(ii)] if $N_H(x) = \{v_1, v_2, v_4, v_5\}$, then  $\{y, x, v_5, v_3, v_4\}$ induces a
banner in $G$, which is a contradiction.
\end{enumerate}
Since the other cases are symmetric, the claim holds.  $\blacklozenge$

\medskip
By Claim \ref{cl:a1p}, we have $A^+ = A_2^+ \cup A_3^+ \cup A_4^+ \cup A_5^+$.

\begin{clm}\label{cl:ap-minus-b1-clique}   $A^+ \setminus B_2$ is a clique. \end{clm}

\no{\it Proof of Claim~\ref{cl:ap-minus-b1-clique}}. Suppose to the contrary that there are non-adjacent vertices $x, y \in A^+ \setminus B_2$. Then by Claims \ref{cl:a1p}, \ref{cl:a2p},  \ref{cl:a3p}, and \ref{cl:a4p}, and by the definition of $A_5^+$, we have $\{v_2, v_3\} \subseteq N_H(x) \cap N_H(y)$. Since $x, y \in A^+\setminus B_2$ and $Q$ is connected, there exists a path $z_0-\cdot\cdot\cdot-z_p$ inside $Q$ such that $xz_0 \in E$ and $yz_p \in E$, and we choose a shortest such path. Suppose that $p = 0$.  We claim that there is no edge between $\{v_1, v_4\}$ and $\{x, y\}$. For suppose on the contrary and without loss of generality that $v_1x \in E$. Then $\{x, y, z_0, v_1, v_3\}$ induces either a $K_{2, 3}$ in $G$ (if $v_1y \in E$), which contradicts Lemma \ref{banner-free--k23-free}, or a banner in $G$ (if $v_1y \notin E$), which is a contradiction. So the claim holds. But, then $\{z_0, x, y, v_2, v_1\}$ induces a banner in $G$, which is a contradiction. Hence $p \geq 1$.  If $p = 1$, then $\{v_2, x, z_0, z_1, y\}$ induces a $C_5$ in $G$,  a contradiction. So, suppose that $p \geq 2$.
 Then since $\{z_2, z_1, z_0, x, v_3, v_4\}$ does not induce a $P_6$ in $G$, $xv_4 \in E$. Again, since $\{z_0, z_1, z_2, y, v_3, v_4\}$ does not induce a $P_6$ in $G$, $yv_4 \in E$. But, then $\{z_0, x, y, v_2, v_4\}$ induces a banner in $G$, which is a contradiction. Thus, Claim \ref{cl:ap-minus-b1-clique} is proved. $\blacklozenge$

 \begin{clm}\label{cl:b2-clique}   $B_{2}$ is a clique. \end{clm}

\no{\it Proof of Claim~\ref{cl:b2-clique}}. Suppose to the contrary that there are non-adjacent vertices $x, y \in B_2$. Since  $Q$ is connected, there exists a path $z_0-\cdot\cdot\cdot-z_p$ inside $Q$ such that $xz_0 \in E$ and $yz_p \in E$, and we choose a shortest such path. Now,  $\{z_0, x, y, v_1, v_2\}$ induces a banner in $G$ (if $p= 0$),  and $\{z_1, z_0, x, v_4, v_3, v_5\}$ induces a $P_6$ in $G$ (if $p \geq 1$), a contradiction. So the claim holds.  $\blacklozenge$

 \begin{clm}\label{cl:ap-clique}   $A^+$ is a clique. \end{clm}

\no{\it Proof of Claim~\ref{cl:ap-clique}}. Suppose to the contrary that there are non-adjacent vertices $x, y \in A^+$. By Claims \ref{cl:ap-minus-b1-clique} and \ref{cl:b2-clique}, we may assume that $x \in A^+ \setminus B_2$ and $y \in B_2$. Since  $Q$ is connected, there exists a path $z_0-\cdot\cdot\cdot-z_p$ inside $Q$ such that $xz_0 \in E$ and $yz_p \in E$, and we choose a shortest such path. Now, if $p \geq 2$, then $\{y, z_p, \ldots, z_0, x, v_3\}$ induces a path $P_t$ ($t\geq 6$) in $G$, a contradiction. So, $p \leq 1$. Suppose that $p = 0$.  Since $\{v_4, x, y, z_0, v_2\}$ does not induce a banner in $G$, $v_4x \notin E$. But now, $\{z_0, x, y, v_3, v_4\}$ induces a $C_5$ in $G$, which is a contradiction. So, $p = 1$. Then $\{v_4, y, z_1, z_0, x\}$ induces a $C_5$ in $G$ (if $xv_4 \in E$), and $\{v_4, y, z_1, z_0, x, v_2\}$ induces a $P_6$ in $G$ (if $xv_4 \notin E$), a contradiction. So the claim holds.  $\blacklozenge$

\medskip
Since $A^+$ is a separator between $H$ and $Q$ in $G$, we obtain that $V(G')\cap A^+$ is a clique separator in $G'$ between $H$ and
$V(G')\cap Q$ (which contains $v$).  This is a contradiction to the
fact that $G'$ is an atom.   This proves Theorem \ref{prime-p6-Banner-c5-free-implies-house-free}. \hfill{\small $\blacksquare$}

Using Theorem \ref{prime-p6-Banner-c5-free-implies-house-free}, we now prove the following:

\begin{thm} \label{complexity-P6-C5-banner}
 The MWIS problem can be solved in $O(n^2m)$-time for ($P_6$, banner, $C_5$)-free graphs.
\end{thm}

\no{\bf Proof}. Let $G$ be a ($P_6$, banner, $C_5$)-free
graph.  First suppose that $G$ is prime.  By
Theorem~\ref{prime-p6-Banner-c5-free-implies-house-free}, every atom of $G$
is nearly house-free.  Since the MWIS problem in ($P_6$, banner, house)-free graphs
can be solved in time $O(nm)$ (by Theorem~\ref{p6-house-banner-free-complexity}), MWIS can be solved in
time $O(n^2m)$ for $G$, by Theorem~\ref{thm:Tar}.  Then the time
complexity is the same when $G$ is not prime, by Theorem
\ref{thm:LM}. \hfill{\small $\blacksquare$}

\subsection{MWIS problem in ($P_6$, banner)-free graphs}

In this section, we show that the MWIS problem can be solved in time $O(n^3m)$ for ($P_6$, banner)-free graphs. In \cite{BKLM}, it was shown that prime atoms of ($P_6$, banner)-free graphs are nearly $C_5$-free. Applying Corollary 9 in \cite{BH} which used an approach for solving MWIS by combining prime graphs and atoms, it was claimed in \cite{BKLM} that MWIS is solvable efficiently in time $O(n^7m)$ for ($P_6$, banner)-free graphs . However, Corollary 9 in \cite{BH} is not proven (and thus has to be avoided); a correct way would be to show that atoms of prime ($P_6$, banner)-free graphs are nearly $C_5$-free (see also \cite{BG, KM} for examples). This will be done in the proof of Theorem \ref{prime-p6-Banner-free-implies-c5-free}.
Though the proof given here is very similar to that of \cite{BKLM}, here we carefully analyze and reprove it so as to apply the known theorems stated in Section~1.

\begin{thm} \label{prime-p6-Banner-free-implies-c5-free}
Let $G$ be a prime ($P_6$, banner)-free graph. Then every
atom of $G$ is nearly $C_5$-free.
\end{thm}

\no {\bf Proof}. Let $G'$ be an atom of $G$.  We want to show
that $G'$ is nearly $C_5$-free, so let us assume on the contrary
that there is a vertex $v \in V(G')$ such that $G'\setminus N[v]$
contains an induced $C_5$, say $H$ with vertices $\{v_1, \ldots, v_5\}$ and edges $v_iv_{i+1}$, for $i \in \{1,  \ldots, 5\}$ ($i$ mod $5$).  For $i=1, \ldots, 5$ we define sets $A_i$,
$A_i^+$, $A_i^-$, $A^+$, $A^-$, and $Q$ as in the last paragraph of
Section~2.

Note that by the definition of $Q$ and $A^+$, we have $A^+ = N(Q)$.  Hence $A^+$ is a separator between $H$ and $Q$ in $G$. Throughout this proof, we take all the
subscripts of $v_i$ to be modulo $5$.  Then we have the following:

Since $G$ is ($P_6$, banner)-free, it is easy to see that $A_1^+ \cup A_2^+ \cup A_4^+ = \emptyset$.  So, $A^+ = A_3^+ \cup A_5^+$.

\begin{clm}\label{cl2:a3}  If $x \in A_3$, then $N_H(x) = \{v_{i-1}, v_i, v_{i+1}\}$, for some  $i\in \{1, \dots, 5\}$.\end{clm}

\no{\it Proof of Claim~\ref{cl2:a3}}. Suppose not. Up to symmetry and without loss of generality, we may assume that $\{v_i, v_{i+2}\} \subseteq N_H(x)$. Now, $\{y, x, v_i, $ $v_{i+1}, v_{i+2}\}$ induces a banner in $G$, which is a contradiction. Hence the claim. $\blacklozenge$

%\begin{clm}\label{cl2:a1p} $A_1^+ = \emptyset$. \end{clm}
%
%\no{\it Proof of Claim~\ref{cl2:a1p}}. Suppose to the contrary that $x \in A_1^+$. Then by definition of $A_1^+$, there exists a vertex $y \in Q$ such that
%$xy \in E$.  Up to symmetry, we may assume that $N_H(x) = \{v_1\}$. Now, $\{y, x, v_1, v_2, v_3, v_4\}$ induces a $P_6$ in $G$, which is a
%contradiction. Thus, Claim \ref{cl2:a1p} is proved. $\blacklozenge$

%\begin{clm}\label{cl2:a2-a3}  (i) If $x \in A_2$, then there exists an $i \in \{1, \dots, 5\}$  such that $N_H(x) = \{v_i, v_{i+1}\}$. (ii) If $x \in A_3$, then there exists an $i\in \{1, \dots, 5\}$ such that $N_H(x) = \{v_{i-1}, v_i, v_{i+1}\}$.\end{clm}
%
%\no{\it Proof of Claim~\ref{cl2:a2-a3}}. Suppose not. Since $x \in A_2 \cup A_3$, up to symmetry and without loss of generality, we may assume that $\{v_i, v_{i+2}\} \subseteq N_H(x)$. Now, $\{y, x, v_i, $ $v_{i+1}, v_{i+2}\}$ induces a banner in $G$, which is a contradiction. Hence the claim. $\blacklozenge$

%\begin{clm}\label{cl2:a2p} $A_2^+ = \emptyset$. \end{clm}
%
%\no{\it Proof of Claim~\ref{cl2:a2p}}. Suppose to the contrary that $x \in A_2^+$. Then by definition of $A_2^+$, there exists a vertex $y \in Q$ such that
%$xy \in E$. Also, since $x \in A_2$, by Claim 2(i), there exists an $i\in \{1, \dots, 5\}$  such that $N_A(x) = \{v_i, v_{i+1}\}$. Up to symmetry, we may assume that $i =1$. Now, $\{y, x, v_2, v_3, v_4, v_5\}$  induces a $P_6$, which is a contradiction. So the claim holds. $\blacklozenge$

\medskip

By Claim \ref{cl2:a3}, define sets $D_{i} = \{x \in A_3^+ \mid N_H(x) = \{v_{i-1}, v_i, v_{i+1}\}\}$, for $i \in \{1, \ldots, 5\}$ ($i$ mod 5).
So, $A_3^+ = \cup_{i =1}^5 D_i$. Now we prove the following: % Then it is easy to see that $D_i$ is a clique, for every $i$, for otherwise, $G$ induces a banner.

\begin{clm}\label{cl2:a3p} $A_3^+$ is a clique.\end{clm}

\no{\it Proof of Claim~\ref{cl2:a3p}}.  Suppose to the contrary that there are non-adjacent vertices $x, y \in A_3^+$. Since $x \in A_3^+$, there exists a vertex $z \in Q$ such that
$xz \in E$. Also,  $x \in D_i$ and $y \in D_j$, for some $i$ and $j$, where $i, j \in \{1, \ldots, 5\}$.  Now:

\begin{enumerate}

\item[(i)] If $x, y \in D_1$, then $\{x, y, v_2, v_3, v_5\}$ induces a banner in $G$, which is a contradiction.

\item[(ii)] If $x \in D_1$ and $y \in D_2$, then since $\{z, x, v_5, v_4, v_3, y\}$ does not induce a $P_6$ in $G$, $yz \in E$. But, then $\{z, x, v_2, y, v_5\}$ induces a banner in $G$, which is a contradiction.

\item[(iii)] If $x \in D_1$ and $y \in D_3$, then since $\{z, x, v_2, y, v_4\}$ does not induce a banner in $G$, $yz \notin E$.
Since $y \in A_3^+$, there exists a vertex $z' \in Q$ such that
$yz' \in E$. Again, since $\{z', x, v_2, y, v_4\}$ does not induce a banner in $G$, $xz' \notin E(G)$. Then since $\{z, x, v_5, v_4, y, z'\}$ does not induce a $P_6$ in $G$, $zz' \in E(G)$. But, now $\{v_1, x, z, z', y, v_4\}$ induces $P_6$ in $G$, which is a contradiction.
\end{enumerate}
Since the other cases are symmetric, Claim \ref{cl2:a3p} is proved. $\blacklozenge$

\begin{clm}\label{cl2-a5p-clique} $A_5^+$ is a clique.\end{clm}

\no{\it Proof of Claim~\ref{cl2-a5p-clique}}. Suppose to the contrary that there are non-adjacent vertices $x, y \in A_5^+$. Since $x \in A_5^+$, there exists a vertex $z \in Q$ such that $xz \in E$. Then since $\{v_1, v_3, x, y, z\}$ does not induce a banner in $G$, $yz \in E$. But, then $\{v_1, v_3, x, y, z\}$ induces a $K_{2, 3}$ in $G$, which is a contradiction to Lemma \ref{banner-free--k23-free}. Thus, Claim \ref{cl2-a5p-clique} is proved. $\blacklozenge$

\begin{clm}\label{cl2-a3p-a5p-complete} $A_3^+$ and $A_5^+$ are complete to each-other.\end{clm}

\no{\it Proof of Claim~\ref{cl2-a3p-a5p-complete}}. Suppose to the contrary that there are non-adjacent vertices $x \in A_3^+$  and $y \in A_5^+$. Up to symmetry, by Claim~\ref{cl2:a3}, we may assume  that $N_H(x) = \{v_1, v_2, v_3\}$.  Since $y \in A_5^+$, there exists a vertex $z \in Q$ such that $yz \in E$. Then since $\{x, v_1, v_3, y, z\}$ does not induce a banner in $G$,  $xz \in E$. But, then $\{z, x, v_1, y, v_4\}$ induces a banner in $G$, which is a contradiction. So the claim holds. $\blacklozenge$

\medskip

Since $A^+ = A_3^+ \cup A_5^+$, and by Claims~\ref{cl2:a3p},  \ref{cl2-a5p-clique} and \ref{cl2-a3p-a5p-complete}, we see that $A^+$ is a clique. Since $A^+$ is a separator between $H$ and $Q$ in $G$, we obtain that $V(G')\cap A^+$ is a clique separator in $G'$ between $H$ and
$V(G')\cap Q$ (which contains $v$).  This is a contradiction to the
fact that $G'$ is an atom.   This proves Theorem \ref{prime-p6-Banner-free-implies-c5-free}. \hfill{$\Box$}

\medskip

Using Theorem \ref{prime-p6-Banner-free-implies-c5-free}, we now prove the following:

\begin{thm}
 The MWIS problem can be solved in $O(n^3m)$-time for ($P_6$, banner)-free graphs.
\end{thm}

\no{\bf Proof}. Let $G$ be an ($P_6$, banner)-free
graph.  First suppose that $G$ is prime.  By
Theorem~\ref{prime-p6-Banner-free-implies-c5-free}, every atom of $G$
is nearly $C_5$-free.  Since the MWIS problem can be solved in time $O(n^2m)$ for ($P_6$, banner, $C_5$)-free graphs
 (by Theorem~\ref{complexity-P6-C5-banner}), MWIS can be solved in
time $O(n^3m)$ for $G$, by Theorem~\ref{thm:Tar}.  Then the time
complexity is the same when $G$ is not prime, by Theorem
\ref{thm:LM}.

\hfill {\small $\blacksquare$}

\medskip
\no{\bf Acknowledgement}: \emph{The author sincerely thanks Prof.Fr\'ed\'eric Maffray for the fruitful discussions.}

\begin{footnotesize}

\end{footnotesize}

\end{document}